\documentclass[final,5p,times,twocolumn,compress,colorlinks = true,linkcolor = blue,urlcolor = blue,citecolor = blue,anchorcolor = blue]{elsarticle}

\usepackage{amssymb}
\usepackage{amsmath}
\usepackage{graphicx}
\usepackage{float}
\usepackage{dcolumn}
\usepackage{multirow}
\usepackage[utf8]{inputenc}
\usepackage[T1]{fontenc}
\usepackage{enumerate}
\usepackage{times}
\usepackage{microtype}
\usepackage{orcidlink}
\usepackage{stfloats}
\usepackage{array}
\usepackage{braket}
\usepackage{subcaption}  
\usepackage{caption}
\usepackage{overpic}

\usepackage{ulem}
\usepackage{xcolor}

\journal{Physics Letters A}

\begin{document}

\begin{frontmatter}



\title{Statistical complexity as a probe of mass and phase structure in compact objects}

\author[first]{Ch.D. Giannios}
\ead{cgiannio@auth.gr}
\author[second,first] {P.S. Koliogiannis\corref{cor1}\orcidlink{0000-0001-9326-7481}}
\ead{pkoliogi@phy.hr}
\author[first]{Ch.C. Moustakidis\orcidlink{0000-0003-3380-5131}}
\ead{moustaki@auth.gr}

\cortext[cor1]{P.S. Koliogiannis}

\affiliation[first]{organization={Department of Theoretical Physics, Aristotle University of Thessaloniki},%Department and Organization
            %addressline={}, 
            city={Thessaloniki},
            postcode={54124}, 
            %state={},
            country={Greece}}
            
\affiliation[second]{organization={Department of Physics, Faculty of Science, University of Zagreb, Bijeni\v cka cesta 32},%Department and Organization
            %addressline={}, 
            city={Zagreb},
            postcode={10000}, 
            %state={},
            country={Croatia}}
            
%% Abstract
\begin{abstract}
In this work, we present a comprehensive and systematic study of the statistical complexity, originally introduced by López-Ruiz, Mancini, and Calbet [Phys. Lett. A 209, 321-326 (1995)], across a broad range of compact star models. We explore how complexity correlates not only with macroscopic observables such as mass and radius, but also with the microscopic characteristics of the underlying equation of state. By incorporating both realistic equations of state and analytical solutions to Einstein’s field equations, we demonstrate that gravitational mass plays a dominant role in determining the behavior of complexity. Furthermore, we show that strong phase transitions within the stellar interior, such as those hypothesized in hybrid stars, can manifest as distinct features in the complexity profile, offering a potential informational signature of such transitions. This work offers new insights into the link between information theory and compact object physics, highlighting complexity's potential as a diagnostic tool in astrophysics.
\end{abstract}


\end{frontmatter}

%============================================================================================================
\section{Introduction}
%============================================================================================================
In recent years, there has been significant interest in studying astrophysical objects using the concept of information entropy and related measures. Notably, Sañudo and Pacheco~\cite{Sanudo-2009} explored the relationship between complexity and the structure of white dwarfs. This approach was later extended to neutron star structures~\cite{Moustakidis-2009}, revealing that the interaction between gravity, short-range nuclear forces, and very short-range weak interactions suggests that neutron stars, within the current theoretical framework, are ordered systems. Similar studies took place in the following years in a series of papers~\cite{de_Avellar-2012,de_Avellar-2014,Adhitya_2020} and Herrera \textit{et al.}~\cite{Herrera-2018a,Herrera-2018b,Herrera-2019a,Herrera-2019b} refined the definition of complexity factors in self-gravitating systems, offering an alternative approach to the problem. Further applications of information measures, in compact objects, can be found in Refs.~\cite{Sharif-2018a,Sharif-2018b,Sharif-2019,Sharif-2022,Yousaf-2020a,Yousaf-2020b,Yousaf-2020c,Yousaf-2021,Yousaf-2022,PhysRevE.66.011102,Bustos_2023,herrera2023,BARGUENO2022169012,Nazar_2025,YE2024101684,Das_2024,Contreras-2021,REHMAN2025116897}. 

In general, a range of complexity measures has been employed to date, each adapted to the specific conditions and constraints of the physical systems under study. These information-theoretic and complexity-based approaches have been applied across a wide array of quantum many-body systems, including nuclei, atoms, atomic clusters, bosons, and molecules~\cite{Adhitya_2020,doi:10.1142/S0218127401003711,LOPEZRUIZ1995321,10.1063/1.2121610,MARTIN2003126,SEN2007286,PANOS200778,MOUSTAKIDIS20181563,LOPEZRUIZ2005215,10.1063/1.1695600,ANTENEODO1996348,BORGOO2007186,10.1063/1.2907743,ANGULO2008670,SANUDO20085283,Sañudo_2008,Patil_2007,10.1063/1.2741244,Garbaczewski_2006,Luzanov10102007,Antolin_2008,Roza_2009,SAGAR200872}.

A specific application of information measures is configurational entropy (CE), introduced by Gleiser and Stamatopoulos~\cite{Gleiser-2012} to explore the link between dynamical and informational aspects of physical models with localized energy configurations. In the following years, CE has been applied in various related studies~\cite{Gleiser-2013,Gleiser-2015a,Gleiser-2015b,Braga-2019,Braga-2020,Yunes-2018,Rocha-2021,Karapetyan-2018,Correa-2016,Baretto-2022,PhysRevD.107.044069,PhysRevD.110.104077,Koliogiannis_RMF}. Gleiser and Jiang~\cite{Gleiser-2015a} explored the link between the stability of compact objects and information-entropic measures. 
They discovered that minimizing CE offers an alternative approach to predicting stability via the maximum mass configuration. Although a formal proof remains elusive, it is intuitively reasonable to expect that CE attains its minimum at this point, as the maximum mass typically represents the most compact and stable configuration. The above statements have been studied in a systematic way recently in Refs.~\cite{PANOS200778,PhysRevD.107.044069,PhysRevD.110.104077,Koliogiannis_RMF}. The main finding supports a strong correlation between the stability points predicted by CE and those obtained through traditional methods, with the accuracy of this correlation showing a slight dependence on the interaction strength. Consequently, the stability of compact objects, composed of components obeying Fermi or boson statistics, can alternatively be assessed using the concept of CE~\cite{PANOS200778,PhysRevD.107.044069,PhysRevD.110.104077,Koliogiannis_RMF}. While CE serves as a possible indicator of the system's stability, another quantity, namely, statistical complexity (SC), offers complementary insight into the internal spatial organization of matter. SC reflects how ordered or structured the density profile is, and thus encodes different aspects of the system's informational content. A combined analysis of CE and SC may therefore provide a more complete characterization of compact objects, offering deeper understanding of both their global stability and internal structural features.

The main motivation of the present work is to extend the study of SC to compact astrophysical systems, including white dwarfs, neutron stars, quark stars, and more generally objects composed of fermionic or bosonic matter, whose structure is characterized by different equations of state (EoSs). In particular, we aim to systematically and comprehensively analyze the SC, as introduced by López-Ruiz, Mancini, and Calbet~\cite{LOPEZRUIZ1995321}, in the aforementioned objects. While several similar studies have been carried out, none have systematically and comprehensively addressed all compact objects, each defined by a distinct EoS~\cite{Sanudo-2009,Moustakidis-2009,de_Avellar-2012,de_Avellar-2014,Adhitya_2020}. Moreover, no other relevant study has emphasized the potential influence of phase transformations on SC values. 

In view of the above, this study aims to establish correlations between SC and both the macroscopic properties of compact objects, such as mass and radius, and the microscopic characteristics of the EoS that govern their internal structure, including the density profile. Special attention is devoted to the role of the crust, with particular emphasis on examining whether potential phase transitions within these stars influence the relationship between SC and mass. Notably, this is the first study in the relevant literature to investigate SC in hybrid stars, which exhibit pronounced phase transformations in their interiors. Furthermore, we seek to investigate whether the SC is specifically connected to the CE in the context of compact object stability. Should such a link exist, it could offer an additional tool for enhancing our understanding of stability from the standpoint of information theory.

The paper is organized as follows: In Sec.~\ref{sec:formalism} we review the theoretical framework for the SC, and in Sec.~\ref{sec:EoS} we present the parametrization for the EoSs. Section~\ref{sec:hydro_equi} is dedicated to the hydrodynamic equilibrium where a set of analytical solutions is also presented, and in Sec.~\ref{sec:results} we discuss the results of the current work. Finally, Sec.~\ref{sec:remarks} contains the scientific remarks of the study.

%============================================================================================================
\section{Theoretical framework for complexity measures} \label{sec:formalism}
%============================================================================================================
To study the SC of a system, we employ a quantity defined by L{\'o}pez-Ruiz, Mancini, and Calbet~\cite{LOPEZRUIZ1995321}, based on an alternative definition proposed in Ref.~\cite{PhysRevE.66.011102}. Specifically, we define the SC as
\begin{flalign}
& C=H\cdot D, \quad \text{with} \quad H=e^{S}, &
\label{eq:complexity}
\end{flalign}
where the quantities $S$ and $D$ represent the Shannon information entropy~\cite{Shannon_1948} and the disequilibrium~\cite{Onicescu_1966}, respectively. The quantity $S$ corresponds to the entropy of a continuous probability distribution $\rho({\bf r})$, which quantifies the uncertainty associated with the distribution, while $D$ measures the disequilibrium, defined as the quadratic distance from equiprobability. It should be noted that the exponential functional preserves the positivity of $C$. The expressions for $S$ and $D$ are given by the following equations
\begin{flalign}
& S[\rho({\bf r})]=-\int \rho({\bf r})\ln \rho({\bf r}) d{\bf r}, \label{eq:c_S}\\ 
& D[\rho({\bf r})]=\int \rho^2({\bf r}) d{\bf r}. &
\label{eq:c_D}
\end{flalign}
It is worth noticing that for a continuous probability distribution, the disequilibrium is the same measure as the information energy defined by Onicescu~\cite{Onicescu_1966}. 

Since this study is focused on compact objects, the definitions of $S$ and $D$ must be accordingly modified to account for the unique properties of these systems. Specifically, the probability distribution is replaced by the mass density $\rho({\bf r})$, which is not derived from the solution of the Schrödinger equation but obtained instead from the solution of the Tolman–Oppenheimer–Volkoff (TOV) equations~\cite{Shapiro-1983,Glendenning-2000,Haensel-2007,Zeldovich-71}, ensuring hydrostatic equilibrium. It is important to emphasize that, although the mass density does not directly correspond to a quantum mechanical probability distribution, it is proportionally related to the probability of finding a particle in a specific location. Hence, it can effectively serve as a suitable probability distribution in this context.

For compact objects, the total mass is given by the integral
\begin{flalign}
& M=\int   \rho({\bf r})  d{\bf r}, \quad \text{or} \quad M/M_{\odot}=b_0 \int \bar{{\cal E}}({\bf r}) d{\bf r}, &
\label{eq:c_mass}
\end{flalign}
where $b_0=8.9\times 10^{-7}$ km$^{-3}$, and we use the relation $\rho({\bf r})={\cal E}({\bf r})/c^2$ for the density, and $\bar{{\cal E}}({\bf r})={\cal E}({\bf r})/\epsilon_0$ for the dimensionless energy density, with $\epsilon_0=1$ MeV fm$^{-3}$. These expressions allow us to calculate the total mass in terms of the energy density within the object~\cite{Moustakidis-2009}.

Utilizing Eq.~\eqref{eq:c_mass}, we redefine the functional forms of Eqs.~\eqref{eq:c_S} and~\eqref{eq:c_D} as
\begin{flalign}
& S[\bar{{\cal E}}({\bf r})]=-b_0\int \bar{{\cal E}}({\bf r})\ln \bar{{\cal E}}({\bf r}) d{\bf r}, & \label{eq:c_S_r}\\
& D[\bar{{\cal E}}({\bf r})]=b_0\int \bar{{\cal E}}^2({\bf r}) d{\bf r}, &
\label{eq:c_D_r}    
\end{flalign}
where $b_0$ is a proper constant chosen to ensure that both the information entropy $S$ and the disequilibrium $D$ are dimensionless quantities, allowing for consistent physical interpretations within the framework of the model.

%============================================================================================================
\section{Equations of state} \label{sec:EoS}
%============================================================================================================
The present study focuses on the behavior of SC and its constituent quantities, namely the Shannon information entropy and the disequilibrium, in various types of compact objects. These include fermionic and bosonic systems (based on Fermi and boson interactions), neutron stars, quark stars, hybrid stars, and white dwarfs, thereby covering a broad range of systems. In the following, we introduce the relevant equations and their parametrizations that are essential for understanding the SC in these different astrophysical contexts.

%============================================================================================================
\subsection{Interacting Fermi gas (FG)}
\label{sec:fermi_gas}
%============================================================================================================
For compact objects composed purely of an interacting Fermi gas (FG), we examined the simplest extension of the free fermion gas by including an additional term that accounts for the repulsive interaction between fermions. Consequently, the energy density and pressure of the fermions are given by the following expressions (for a detailed analysis, see Ref.~\cite{Narain-06}) 
~
\begin{flalign}
& {\cal E}(n_{\chi})=\frac{(m_{\chi}c^2)^4}{(\hbar c)^38\pi^2}\left[x\sqrt{1+x^2}(1+2x^2) \right. \nonumber\\
&\quad\quad\quad -\left. \ln(x+\sqrt{1+x^2})\right] + \frac{y^2}{2} (\hbar c)^3 n_{\chi}^2, \\
& P(n_{\chi})=\frac{(m_{\chi}c^2)^4}{(\hbar c)^38\pi^2}\left[x\sqrt{1+x^2}(2x^2/3-1)
\right. \nonumber \\
&\quad\quad\quad +\left. \ln(x+\sqrt{1+x^2})\right]  + \frac{y^2}{2} (\hbar c)^3 n_{\chi}^2, &
\label{Rel-ED}
\end{flalign}
where $m_{\chi}$ is the particle mass, considered equal to $m_{\chi}=939.565~{\rm MeV/c^{2}}$ for reasons of simplicity (this holds throughout the study), $n_{\chi}$ is the number density of the fermions, and 
\begin{flalign}
   & x=\frac{(\hbar c)(3\pi^2n_{\chi})^{1/3}}{m_{\chi}c^2}. &
   \label{eq:chi_quantity}
\end{flalign}
The parameter, $y$ (in units of $\rm MeV^{-1}$), is the one that introduces the repulsive interaction.
An increase in $y$
corresponds to a stronger interaction  and vice-versa. The corresponding values are shown in Table~\ref{tab:interaction_parameters}.

\begin{table}
	\caption{Interaction parameters for the FG and BG cases. $y$ is in units of $\rm MeV^{-1}$, and $w$ and $z$ in units of $\rm MeV^{-1}~fm^{3}$.}
    \begin{center}
    \footnotesize
		\begin{tabular}{l c c c c c c c c}
            \hline
            \hline
            \multicolumn{9}{c}{\vspace{-0.25cm}} \\
            \vspace{0.15cm} FG & $y$ & 0 & -- & -- & 0.05 & 0.1 & 0.3 & 0.5 \\
            \vspace{0.1cm} \multirow[t]{3}{*}{BG} & $w$ & -- & 0.005 & 0.01 & 0.05 & -- & -- & 0.5 \\
            \vspace{0.1cm}  & $z$ & -- & 0.005 & 0.01 & 0.05 & -- & -- & 0.5 \\
            \hline
		\end{tabular}
   \end{center}
	\label{tab:interaction_parameters}
\end{table}

%===========================================================================================================
\subsection{Interacting boson gas (BG)}
\label{sec:boson_gas}
%===============================================================%============================================
To broaden our analysis, we also considered compact stars composed of bosonic matter, specifically focusing on an interacting boson gas~\cite{JETZER1992163,Liebling_2012,doi:10.1142/S0218271821300068}. Since the construction of the EoS for the aforementioned gas is not uniquely defined, we introduce two distinct cases based on varying assumptions. 

\begin{enumerate}
    \item \textbf{BG-C1}: The first derivation of the EoS for boson stars with a repulsive interaction was presented in Refs.~\cite{Colpi-1986,PhysRevD.105.023001,PhysRevD.109.043029}. Since then, it has been widely used in corresponding calculations. In particular, the energy density is expressed as follows
    \begin{flalign}
    & {\cal E}(P)= \frac{4}{3w} \left[ \left(\sqrt{\frac{9w}{4}P}+1  \right)^2-1 \right], &
    \label{eq:boson_matter_a}
    \end{flalign}
    where $w=4\lambda (\hbar c)^3/(m_{\chi}c^2)^4$ (in units of $\rm MeV^{-1}~fm^{3}$). In fact, the parameter 
    $\lambda$ is associated with the strength of the interaction. However, it is usual to employ the combination of $m_{\chi}$ and $\lambda$ defined as $w$. 
    An increase in $w$ enhances the strength of the interaction and vice-versa. The corresponding values are shown in Table~\ref{tab:interaction_parameters}.

    \item \textbf{BG-C2}: The second approach to describing the interior of a boson star is through the EoS given in Ref.~\cite{Agnihotri-2009} and used recently in Ref.~\cite{Watts-2023}, where the energy density is given by
    \begin{flalign}
    & {\cal E}(P)=P +\sqrt{\frac{2P}{z}}, &
    \label{eq:boson_matter_b}
    \end{flalign}
    with the interaction parameter $z$ being equal to $z=u^2(\hbar c)^3/(m_{\chi} c^2)^2$ (in units of $\rm MeV^{-1}~fm^{3}$) and the quantity $u={\rm g}_{\chi}/m_{\phi}c^2$ defining the strength of the interaction in analogy to the case of fermions. 
    The corresponding values are shown in Table~\ref{tab:interaction_parameters}.

\end{enumerate}

%============================================================================================================
\subsection{Hadronic EoSs}
\label{sec: Hadron-EoS}
%============================================================================================================
For the description of neutron stars, we employ a set of hadronic EoSs that have been widely used in the literature for studies of neutron star properties (see Ref.~\cite{Koliogiannis-2020} and references therein). The above EoSs are constructed to describe the inner shell of neutron stars, that is, the core, which is a quantum fluid composed of nucleons (mainly protons and neutrons) and electrons. In each case, the region of the star's crust, which is in a crystalline state, is described by the EoS of Feynman, Metropolis and Teller~\cite{PhysRev.75.1561}, and also of Baym, Bethe, and Sutherland~\cite{Baym-71}. The overall EoS of a neutron star results from the combination of the EoSs describing its core and crust. These EoSs have been selected to ensure that they align with the predicted maximum observed neutron star mass, which falls within the range of $\sim 2 M_{\odot}$, or higher.
%============================================================================================================
\subsection{Quark EoSs}
\label{sec: Quark-EoS}
%============================================================================================================
For the description of quark stars, we use a set of EoSs for interacting quark matter, as predicted and applied in Ref.~\cite{Zhang-2021} (for recent  applications see also Refs.~\cite{SINGH2021100774,BANERJEE2021100792}). In this case, the pressure is connected to the energy density through the simplified expression 
\begin{flalign}
    & \frac{P}{4 B_{\rm eff} }=\frac{1}{3}\left( \frac{\cal E}{4 B_{\rm eff} }-1  \right) \nonumber \\
 &\quad\quad\quad +
 \frac{4}{9\pi^2}\lambda\left( -1+\sqrt{1+\frac{3\pi^2}{\lambda}\left( \frac{\cal E}{4 B_{\rm eff} }-\frac{1}{4} \right)  }  \right). &
\end{flalign}
Specifically, in the present work, we have used the value of $B_{\rm eff}=150~{\rm MeV~fm^{-3}}$, while for the dimensionless parameter $\lambda$, we applied the values of $(1,2,5,16)$~\cite{Zhang-2021}.

%============================================================================================================
\subsection{Hybrid EoSs}
\label{sec: Hybrid-EoS}
%============================================================================================================
To describe hybrid stars, we use a more advanced EoS crafted to model the features of a third family of compact objects, referred to as twin stars~\cite{Bielich-2020,Alford-2013}. Specifically, we employ the Maxwell construction, which is particularly effective for describing phase transitions within a compact object and is formulated as follows~\cite{Alford-2013}
\begin{flalign}
  & \mathcal{E}(P) = \begin{cases} 
      \mathcal{E}_{\rm H}(P), & P\leq P_{\rm tr} \\
      \mathcal{E}_{\rm H}(P_{{\rm tr}}) + \Delta \mathcal{E} + c_s^{-2}(P-P_{{\rm tr}}), & P \geq P_{{\rm tr}}
   \end{cases} &
   \label{MC-1}
\end{flalign}
where $c_{\rm s}=\sqrt{{\partial P}/{\partial {\cal E}}}$ is the speed of sound (in units of the speed of light), and $\Delta \mathcal{E}$ is the magnitude of the energy density jump at the transition point. The subscript $``{\rm tr}"$ denotes the corresponding quantity at this point. In the region $P\leq P_{\rm tr}$,  we utilized the GRDF-DD2 EoS~\cite{Typel-2018,Tsaloukidis-2023} while in the region $P\geq P_{\rm tr}$, the value of the speed of sound is fixed at $c_{\rm s}=1$. 

%============================================================================================================
\subsection{White dwarfs}
%============================================================================================================
To describe white dwarfs, we consider that the energy density arises from both the rest mass of the ions and the energy density of the electrons, with the electrons contributing solely to the pressure~\cite{Shapiro-1983,Glendenning-2000,Haensel-2007}. The total energy density is then given by
\begin{flalign}
 & {\cal E}_{\rm tot}(n_e)={\cal E}_{\rm ions}(n_e)+{\cal E}_{\rm el}(n_e), &
 \label{eq:en_wd}
\end{flalign}
where
\begin{flalign}
 & {\cal E}_{\rm ions}(n_e)=n_em_N c^2\frac{A}{Z}, &
 \label{eq:en_ions_wd}
\end{flalign}
and
\begin{flalign}
& {\cal E}_{\rm el}(n_{e})=\frac{(m_ec^2)^4}{(\hbar c)^38\pi^2}\left[x\sqrt{1+x^2}(1+2x^2) \right. \nonumber\\
&\quad\quad\quad-\left. \ln(x+\sqrt{1+x^2})\right]. &
\label{eq:en_el_wd}
\end{flalign}
In Eqs.~\eqref{eq:en_wd}-\eqref{eq:en_el_wd} $Z$ and $A$ are the atomic and mass numbers of the ions, and $m_N$ and $m_e$ denote the nucleon and electron mass, respectively. In the present work we consider that $A/Z=2$~\cite{Shapiro-1983,Glendenning-2000,Haensel-2007}. The quantity $x$ is defined in the same way as Eq.~\eqref{eq:chi_quantity}, where $n_{\chi}$ and $m_{\chi}$ are replaced by $n_{e}$ and $m_{e}$, respectively. The total pressure, which is only due to electrons, is given by
\begin{flalign}
 & P_{\rm tot}(n_e)\equiv P_{\rm el}(n_e)=\frac{(m_ec^2)^4}{(\hbar c)^38\pi^2}\left[x\sqrt{1+x^2}(2x^2/3-1)
\right. \nonumber \\
&\quad\quad\quad\quad\quad\quad\quad\quad +\left. \ln(x+\sqrt{1+x^2})\right]. &
\label{eq:pr_wd}
\end{flalign}
Moreover, we also consider the so-called non-relativistic case ($x\ll 1$), where the electrons are treated as non-relativistic at low densities. In this regime, the EoS is given by~\cite{Shapiro-1983,Glendenning-2000,Haensel-2007}
\begin{flalign}
& P=K_{\rm non-rel}{\cal E}^{5/3}, \quad K_{\rm non-rel}=\frac{(\hbar c)^2}{15 \pi^2m_e c^2}\left(\frac{3\pi^2 Z}{A m_n c^2}  \right)^{5/3}. &
\label{eq:wd-non-rel}
\end{flalign}
While this equation is valid only for low densities, and its extrapolation to high densities leads to non-physical behavior, we employ it across the full density regime for comparison purposes.

%============================================================================================================
\section{Hydrodynamic equilibrium and analytical solutions} \label{sec:hydro_equi}
%============================================================================================================
To obtain the microscopic and macroscopic properties of compact objects, namely the density distribution and the gravitational mass, which are essential for the calculation of $S$ and $D$, and as a consequence $C$, we use Einstein's field equations for a spherical fluid. In particular, the mechanical equilibrium of the star's matter is governed by the well-known TOV equations~\cite{Shapiro-1983,Glendenning-2000,Haensel-2007,Zeldovich-71}
\begin{flalign}
	& \frac{dP(r)}{dr}=-\frac{G\mathcal{E}(r) M(r)}{c^{2}r^2}\left(1+\frac{P(r)}{\mathcal{E}(r)}\right)\left(1+\frac{4\pi P(r) r^3}{M(r)c^2}\right) \\\nonumber & \quad\quad\quad\quad \left(1-\frac{2GM(r)}{c^2r}\right)^{-1}, \\
    & \frac{dM(r)}{dr}=\frac{4\pi r^2}{c^{2}}\mathcal{E}(r). &
\end{flalign}

Generally, to obtain realistic solutions, the most natural approach is to numerically solve the TOV equations while incorporating an EoS that defines the relationship between pressure and energy density within the fluid interior.  Alternatively, one can attempt to find analytical solutions to the TOV equations, although these solutions may lack physical relevance. Although there are numerous analytical solutions to the TOV equations~\cite{Kramer-1980,Delgaty-1998}, only a few hold significant physical relevance. In this work, we utilize two of these notable solutions: the Schwarzschild solution (constant-density interior) and the Tolman-VII solution~\cite{Kramer-1980,Delgaty-1998}. It is important to note that analytical solutions are highly valuable, as they often provide explicit expressions for key quantities and play a crucial role in verifying the accuracy of numerical calculations.  Below, we provide a brief overview of these two fundamental analytical solutions.

\begin{itemize}
    \item \textbf{Schwarzschild  solution}: For the Schwarzschild interior solution, the density is uniform throughout the star~\cite{Weinberg-72,Schutz-85}. The energy density and pressure are given by
    \begin{flalign}
        & {\cal E}={\cal E }_c=\frac{3Mc^2}{4\pi R^3}, \\
        & \frac{P(x)}{{\cal E}_c }=\frac{\sqrt{1-2\beta}-\sqrt{1-2\beta x^2}}{\sqrt{1-2\beta x^2}-3\sqrt{1-2\beta }}, &
        \label{Unif-E}
    \end{flalign}
    where $x=r/R$, $\beta=GM/Rc^2$ is the compactness of the star, and  ${\cal E}_{c}=\rho_c c^2$ is the central energy density. 

    \item \textbf{Tolman-VII solution}: The Tolman-VII solution has been widely used in neutron star studies, and its physical realization has been recently examined in detail~\cite{Oppenheimer-39,Raghoonundun-2015}. The stability of this solution was examined by Negi {\it et al}.~\cite{Negi-1999,Negi-2001} and also confirmed in Refs.~\cite{Moustakidis-2017,Posada-2021}. The energy density and the pressure read as (for all the mathematical details see Ref.~\cite{Moustakidis-2017})
    \begin{flalign}
        & \frac{{\cal E} (x)}{{\cal E}_c}=(1-x^2), \quad {\cal E}_c=\frac{15Mc^2}{8\pi R^3},\\
        & \frac{P(x)}{{\cal E}_c}=\frac{2}{15}\sqrt{\frac{3e^{-\lambda}}{\beta }}\tan\phi-\frac{1}{3}+\frac{x^2}{5}. &
        \label{Tolm-E}
    \end{flalign}
\end{itemize}
It is worth mentioning that these solutions are applicable to any type of compact object, regardless of its mass and radius. Consequently, they are suitable for studying any type of massive or supramassive object for which hydrodynamic stability is governed by the TOV equations. Regardless, valuable insights can be gained from analytical solutions that examine both the qualitative and quantitative behavior of SC as a function of the bulk properties (mass and radius) of compact objects.

%============================================================================================================
\section{Results and discussion} \label{sec:results}
%============================================================================================================
\subsection{Complexity in Schwarzschild   and Tolman-VII solutions}
%============================================================================================================

\begin{figure*}[t!]
\centering
\includegraphics[width=0.9\textwidth]{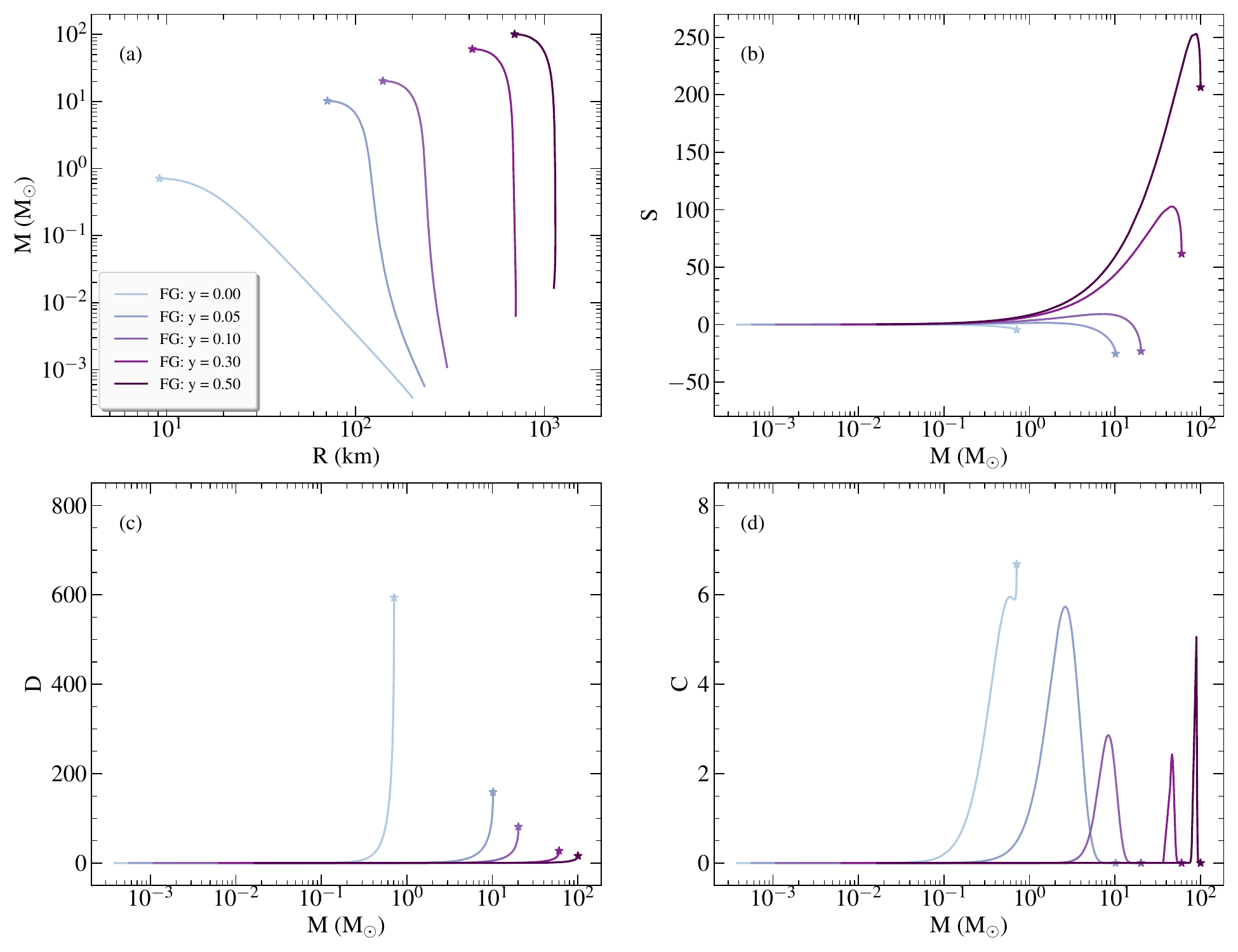}
\caption{(a) Gravitational mass as a function of the radius for a set of FG EoSs. The corresponding dependence of (b) the Shannon information entropy, (c) the disequilibrium, and (d) the SC as a function of the gravitational mass. Stars indicate the stability points. In (d), the curves corresponding to interaction strengths $y=[0.1,0.3,0.5]~{\rm MeV^{-1}}$ have been normalized by the order of magnitude of the corresponding peak, namely $10^{4}$, $10^{45}$, and $10^{110}$.}
\label{fig:FG}
\end{figure*}

\begin{itemize}
    \item {\bf Schwarzschild   solution:}
    The Shannon information entropy, disequilibrium, and SC in the case of the Schwarzschild  solution are derived from Eqs.~\eqref{eq:complexity}-\eqref{eq:c_D} and are given respectively by the following analytical expressions
    \begin{flalign}
        & S_S = \frac{-b_0 M c^2}{\epsilon_0} \ln\left( \frac{3 M c^2}{4 \pi \epsilon_0 R^3} \right), \quad D_S = \frac{3 b_0 M^2 c^4}{4 \pi \epsilon_0^2 R^3},& \\
        & C_S = D_{S}\exp[S_{S}] =\alpha \delta^{1-\alpha\xi} \xi^{2-\alpha \xi}\bar{R}^{3(\alpha \xi - 1)}, &
        \label{eq:schwarzchild_sol_b}
    \end{flalign}
    with $\xi=M/M_{\odot}$ and $\bar{R}=R/{\rm km}$. The dimensionless constants $\alpha$ and $\delta$ are given by
    \begin{flalign}
        & \alpha=\frac{b_0 M_{\odot}c^2}{\epsilon_0}=1,\qquad \delta=\frac{3 M_{\odot}c^2}{4\pi \epsilon_0  {\rm km}^3}=2.685 \times  10^5. \nonumber &
    \end{flalign}
    The value of the mass $M$ and radius $R$ that maximizes the SC results from
    \begin{flalign}
        & \frac{dC_{S}}{dM} = 1 - 6.75\xi - 0.5\xi\ln\left(\xi \bar{R}^{-3}\right) = 0. & 
        \label{eq:schwarzchild_sol}
    \end{flalign}
    In this study, we focus on the specific case where the radius is $R=10~{\rm km}$ (a typical value of radius for neutron stars) and the parameter $\xi=0.359$, which is derived from Eq.~\eqref{eq:schwarzchild_sol}.

    \item {\bf Tolman-VII solution:} In this case the corresponding quantities  are given by
    \begin{flalign}
        & S_T=-\xi
        \left[\ln\left(6.7\times 10^5\xi \bar{R}^{-3}\right)-0.68\right], & \\
        & D_T=3.84\times 10^5\xi^2 \bar{R}^{-3}, & \\
        & C_T = 3.84 \times 10^5\xi^2 \bar{R}^{-3}\exp\left\{ -\xi \left[ \ln\left( \xi \bar{R}^{-3} \right) + 12.735 \right] \right\}. &
        \label{eq:tolman_sol_b}
    \end{flalign}
    Maximizing SC originates through the equation
    \begin{flalign}
        & \frac{dC_{T}}{dM} = 0.766 - 5.26\xi - 0.383\xi\ln\left(\xi \bar{R}^{-3}\right) = 0. &
        \label{eq:tolman_sol}
    \end{flalign}
    Similar to the Schwarzschild solution, in this case, we focus on the specific case where the radius is $R=10~{\rm km}$ and the parameter $\xi=0.347$, which is derived from Eq.~\eqref{eq:tolman_sol}.
    The above results (considering both analytical solutions) are compared with the corresponding ones originated from the solution of TOV equations using realistic EoSs. 

\end{itemize}

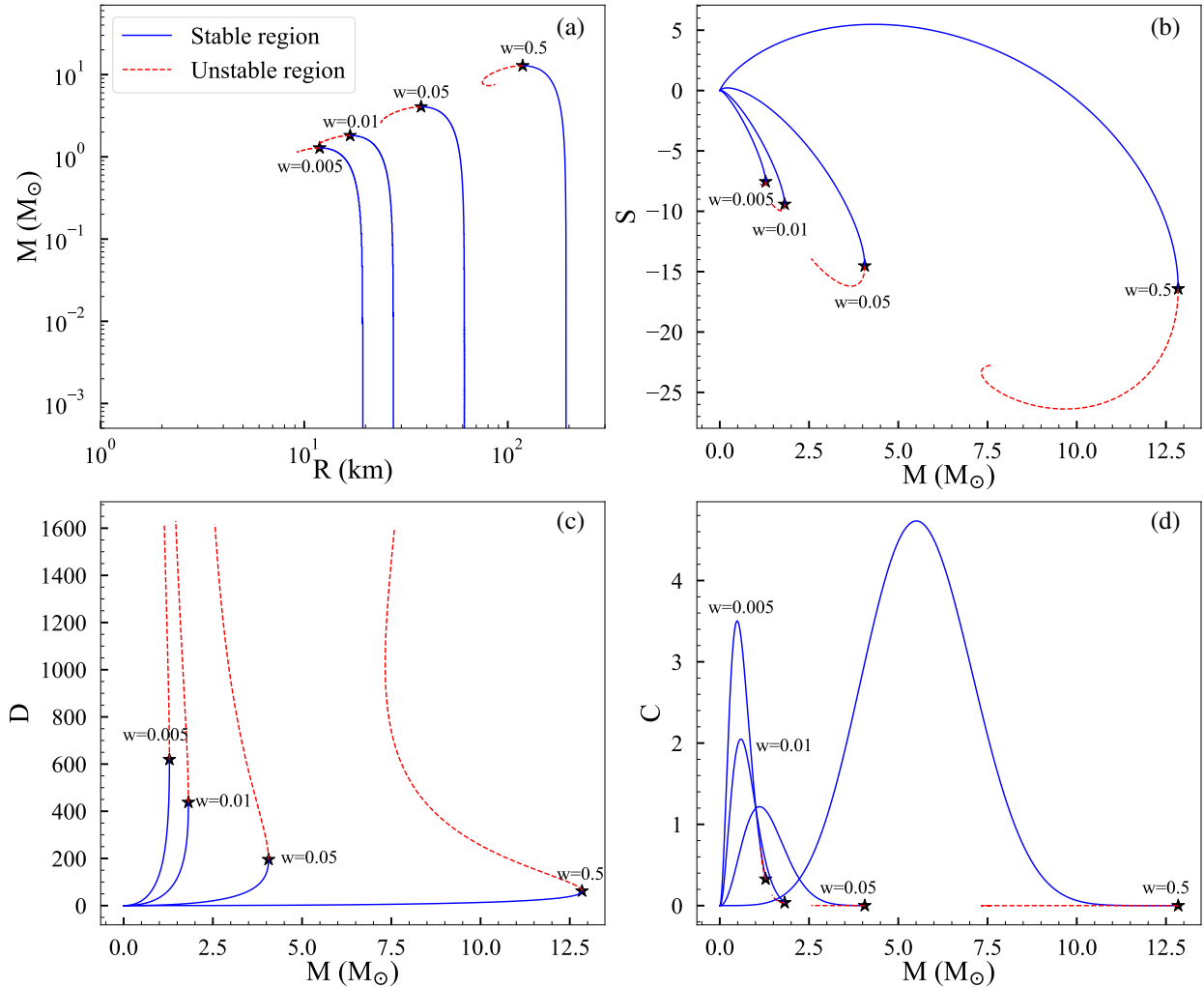
\begin{figure*}[t!]
    \centering
    \begin{overpic}[width=0.9\textwidth]{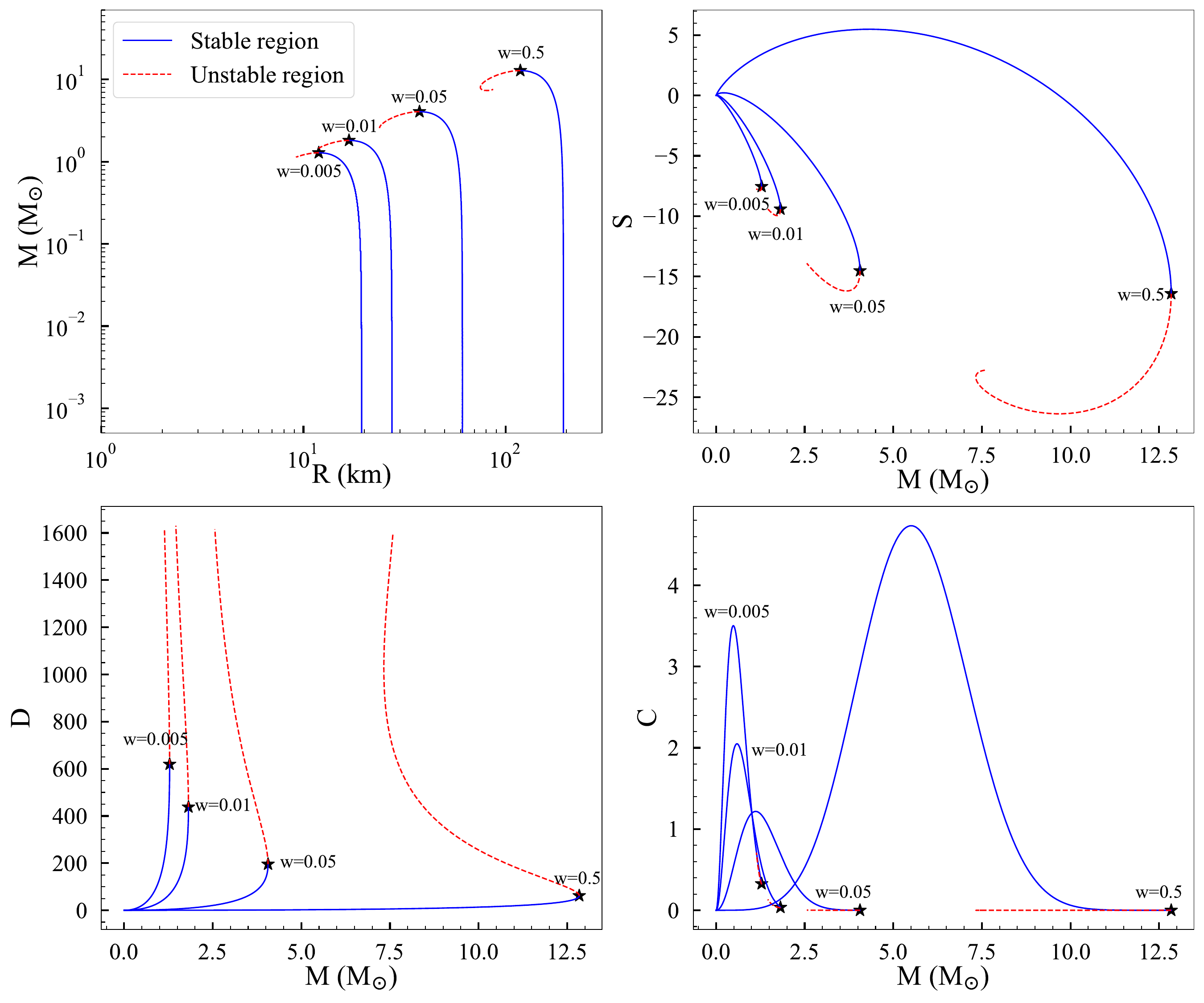}
    \put(46,80){(a)}
    \put(95,80){(b)}
    \put(46,39){(c)}
    \put(95,39){(d)}
    \end{overpic}
    \caption{(a) Gravitational mass as a function of the radius for a set of BG-C1 EoSs. The corresponding dependence of (b) the Shannon information entropy, (c) the disequilibrium, and (d) the SC as a function of the gravitational mass. The stars indicate the stability points. In (d), the curve corresponding to interaction strength $w=0.5~{\rm MeV^{-1}~fm^{3}}$ has been normalized by $10^{2}$.}
    \label{fig:eos_BG_C1}
\end{figure*}

%============================================================================================================
\subsection{Fermion stars}
%============================================================================================================
In Fig.~\ref{fig:FG}, we present various information quantities for the case of the interacting Fermi gas. Specifically, Fig.~\ref{fig:FG}(a) shows the mass-radius dependence, while Figs.~\ref{fig:FG}(b-d) illustrate the Shannon information entropy, disequilibrium, and SC as functions of the compact object's mass, respectively. Stars denote the turning point. Increasing the interaction strength (parameter $y$) results in compact objects with higher values for both maximum mass and the corresponding radii, a trend that is also reflected in the values of $S$, $D$, and $C$. Furthermore, the behavior of the constituent quantities of $C$, $S$ and $D$, leads to the maximization of SC within the stability region. As the interaction strength increases, this maximum is shifted toward higher masses.

In the case of FG EoSs that include interactions, a minimization in $C$ is observed as the stellar mass approaches its maximum value, corresponding to the stability point. However, this trend is not present in the non-interacting FG EoS ($y=0$). For $y=0$, although $C$ reaches a local peak within the stability region, it subsequently increases monotonically. This behavior is closely tied to the mass-radius relation, which exhibits an extended plateau, indicating nearly constant mass over a broad range of radii.
In particular, this behavior can be quantitatively described using the analytical Schwarzschild solution. As indicated by relation~(\ref{eq:schwarzchild_sol_b}),
the complexity behaves as
$C \sim M \bar{\rho}^{1-M}$, where $\bar{\rho}$ is the mean density of the object ($\bar{\rho}\sim M/R^3$).
Consequently,  an increase in mass, provided the mass remains less than 
$1 \ M_{\odot}$, 
results in a corresponding rise in complexity. Since this inequality holds universally for the case 
$y=0$, complexity will continue to grow steadily with mass, without reaching a maximum. Overall, we can conclude that the dependence of complexity on mass stems from the specific form of the mass–radius relationship (M–R diagram), which, in turn, is governed by the type of EoS applied in each case. Similar conclusions arise in the other cases examined (see below).

\begin{figure*}[t!]
    \centering
    \begin{overpic}[width=0.9\textwidth]{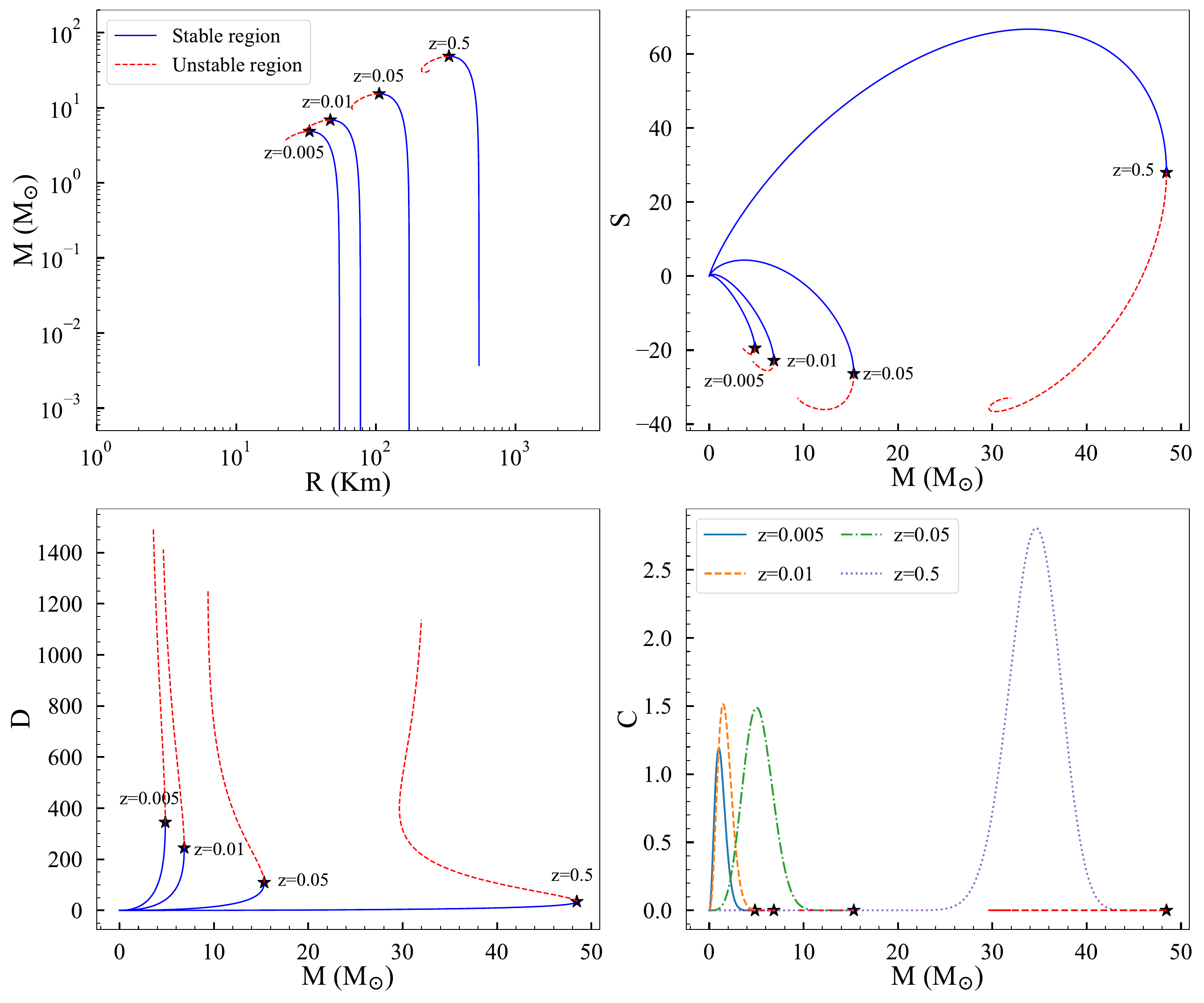}
    \put(46,80){(a)}
    \put(95,80){(b)}
    \put(46,39){(c)}
    \put(95,39){(d)}
    \end{overpic}
    \caption{(a) Gravitational mass as a function of the radius for a set of BG-C2 EoSs. The corresponding dependence of (b) the Shannon information entropy, (c) the disequilibrium, and (d) the SC as a function of the gravitational mass. The stars indicate the stability points. In (d), the curves corresponding to interaction strengths $w=[0.05,0.5]~{\rm MeV^{-1}~fm^{3}}$ have been normalized by $10^{2}$ and $2\times 10^{29}$, respectively.}
    \label{fig:eos_BG_C2}
\end{figure*}

\begin{figure*}[t!]
    \centering
    \begin{overpic}[width=0.9\textwidth]{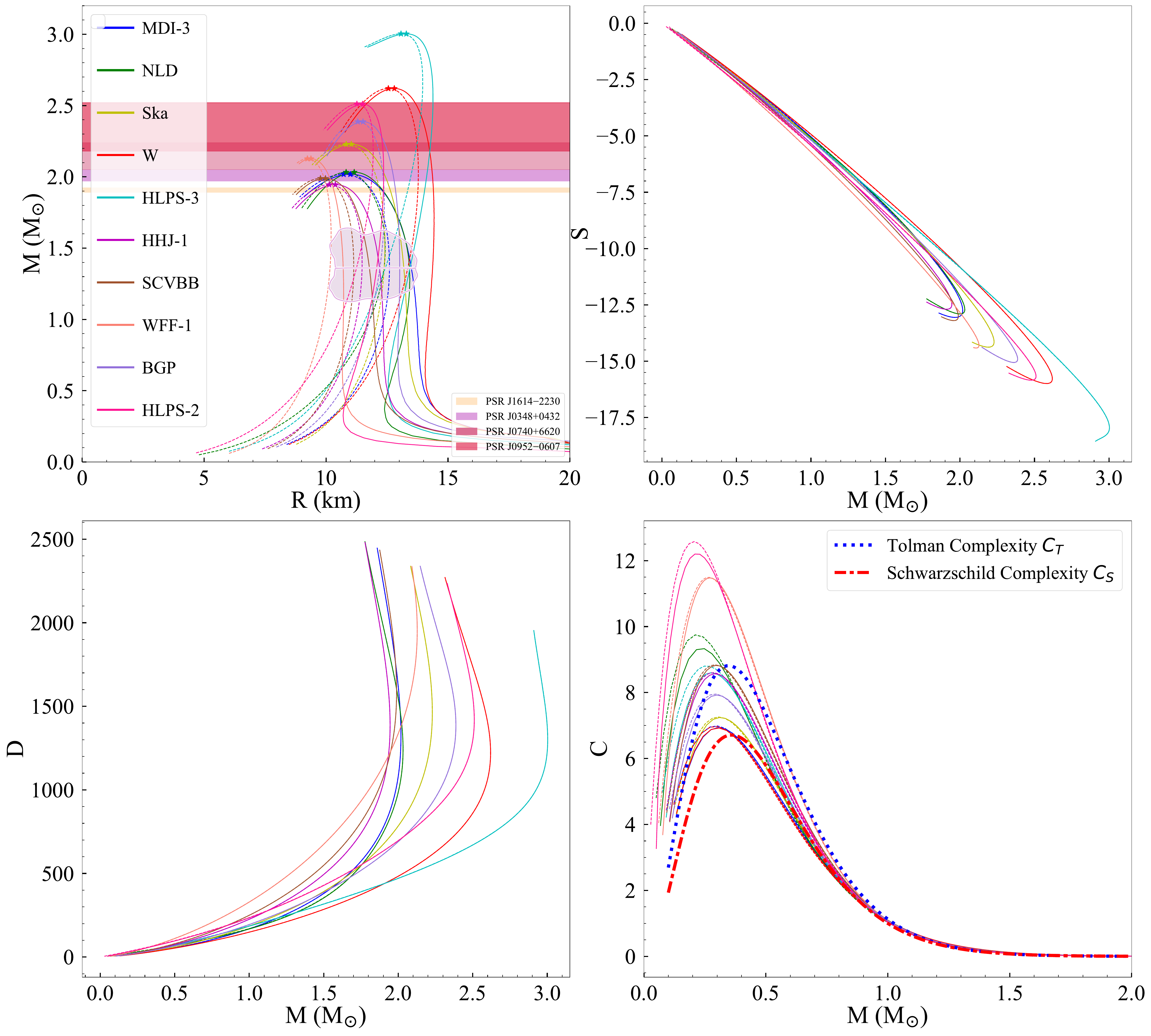}
    \put(46,82){(a)}
    \put(95,82){(b)}
    \put(46,36){(c)}
    \put(95,36){(d)}
    \end{overpic}
    \caption{(a) Gravitational mass as a function of the radius for a set of hadronic EoSs. The corresponding dependence of (b) the Shannon information entropy, (c) the disequilibrium, and (d) the SC as a function of the gravitational mass. The stars indicate the stability points. In (d), the dashed-dotted and dotted curves correspond to Schwarzschild and Tolman-VII analytical solutions, respectively, for a fixed value of radius, $R=10~{\rm km}$. The shaded regions from bottom to top represent the GW170817 event~\cite{Abbott-2019}, and the PSR J1614-2230~\cite{Arzoumanian-2018}, PSR J0348+0432~\cite{Antoniadis-2013}, PSR J0740+6620~\cite{Cromartie-2020}, and PSR J0952-0607~\cite{Romani-2022} pulsar observations with possible maximum neutron star mass.}
    \label{fig:eos_hadrons}
\end{figure*}

%============================================================================================================
\subsection{Boson  stars}
%============================================================================================================
Figures~\ref{fig:eos_BG_C1} and ~\ref{fig:eos_BG_C2} present, in the same manner as Fig.~\ref{fig:FG}, the various informational quantities for the interacting boson gas in the BG-C1 and BG-C2 cases. In all cases, the solid line represents a stable configuration, while the dashed line indicates an unstable one. The star is denoting the turning point. In these cases, the SC is also maximized at a specific value of the gravitational mass within the stable regime of the compact object. Notably, in BG-C1, the maximum SC decreases with increasing interaction strength (parameter $w$), but this trend reverses at higher interaction values ($w=0.5~{\rm MeV^{-1}~fm^{3}}$), a behavior not observed in the BG-C2 EoSs. It is evident that the choice of parametrization for the boson gas EoS has a significant impact on the behavior of the information measures, influencing the relationship between the interaction strength and the corresponding values of SC, Shannon information entropy, and disequilibrium.

By completing the systematic study of stars composed of interacting fermionic and bosonic gases, it is crucial to emphasize that the results indicate no correlation between the maximization of the SC and the maximum mass values corresponding to points of  stability. While previous studies~\cite{Gleiser-2015a,PhysRevD.107.044069,PhysRevD.110.104077,Koliogiannis_RMF} have demonstrated a relationship between CE and stability thresholds, the absence of such correlation in the case of SC suggests that SC is not directly linked to CE. This comparison strengthens the conclusion that SC is primarily governed by the bulk properties of compact objects, showing limited sensitivity to variations in their internal density distributions. In contrast, CE is evidently highly responsive to changes in the internal density profile, thereby serving as a more effective indicator for identifying stability points through its minimization.

%============================================================================================================
\subsection{Hadronic  stars}
%============================================================================================================
Following the analysis of stars based on Fermi and boson interacting gases, we now focus on the case of hadronic stars. It is important to note that this case pertains to neutron stars, where the EoS is formulated in a sophisticated manner, accounting for the presence of not only neutrons, protons, and electrons, but also other particles such as hyperons, muons, and others. We consider this case to be the most realistic, as it corresponds to observed compact objects. Figure~\ref{fig:eos_hadrons} displays 10 representative, realistic EoSs, including the mass-radius dependence and various informational quantities. To provide a complete picture, we overlay in shaded regions key observational constraints on the neutron star mass and radius. From bottom to top, these include bounds from the GW170817 event~\cite{Abbott-2019}, as well as mass measurements of the pulsars PSR J1614-2230~\cite{Arzoumanian-2018}, PSR J0348+0432~\cite{Antoniadis-2013}, PSR J0740+6620~\cite{Cromartie-2020}, and PSR J0952-0607~\cite{Romani-2022}, all of which inform the maximum mass a neutron star can support. Specifically, for each EoS, two parametrizations are considered: (a) one where the EoS of the neutron star's interior is extended to the surface (i.e. no crust), and (b) one where the EoS for the outer crust is incorporated into the EoS of the neutron star's interior. Both parametrizations lead to a striking feature: in all cases, the same qualitative behavior of the information quantities is observed. At this stage, the presence or absence of the crust appears to have minimal impact on the behavior of these quantities. 
\begin{figure*}[t!]
\centering
\begin{overpic}[width=0.9\textwidth]{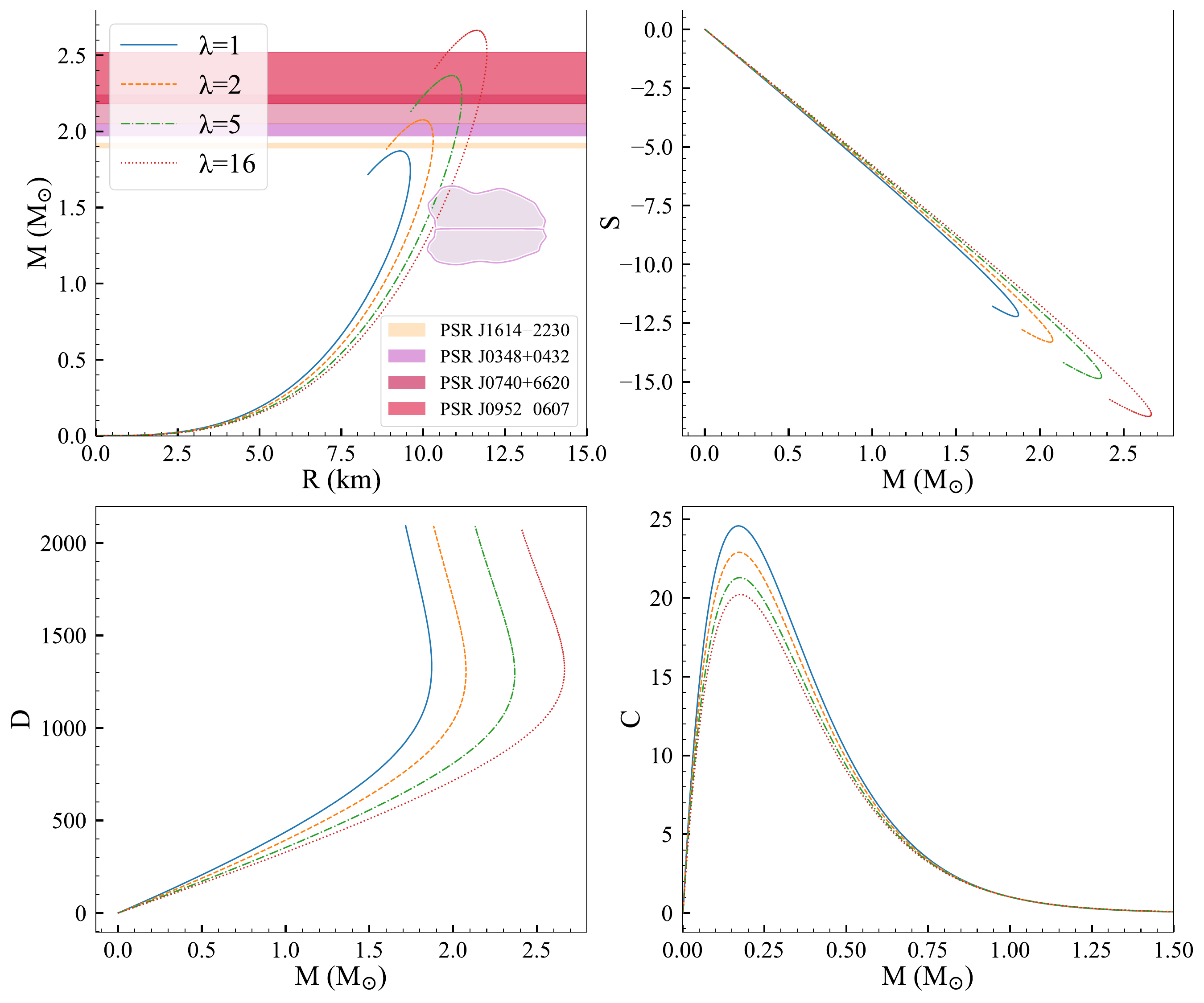}
\put(10,52){(a)}
    \put(59,52){(b)}
    \put(10,10.7){(c)}
    \put(59,10.7){(d)}
    \end{overpic}
\caption{(a) Gravitational mass as a function of the radius for a set of quark EoSs. The corresponding dependence of (b) the Shannon information entropy, (c) the disequilibrium, and (d) the SC as a function of the gravitational mass. The shaded regions from bottom to top represent the GW170817 event~\cite{Abbott-2019}, and the PSR J1614-2230~\cite{Arzoumanian-2018}, PSR J0348+0432~\cite{Antoniadis-2013}, PSR J0740+6620~\cite{Cromartie-2020}, and PSR J0952-0607~\cite{Romani-2022} pulsar observations with possible maximum neutron star mass.}
\label{fig:eos_quarks}
\end{figure*}
Furthermore, in all cases, SC reaches its peak within the mass range of 0.25–0.35 $M_{\odot}$, with the maximum value falling between SC values of 6.5-12.5. It is also noteworthy that all EoSs rapidly converge in SC as mass increases, becoming nearly identical for $M> 1 M_{\odot}$.\\
\indent One can speculate here  that SC remains low at both very low and very high masses due to the dominance of the crustal state in the former and the perfect fluid state in the latter. At intermediate masses, the balanced contribution of these two states results in a more complex system structure.
The aforementioned estimate clearly applies to the case in which the star's structure is described in the most physically realistic manner.
It is worth mentioning that in Ref.~\cite{Adhitya_2020}, the authors, who studied the case of SC in neutron stars with a crust and a hyperon core, concluded that the crust EoS sets an upper limit to the SC value, and this implies that the crust itself constitutes an ordered system with low SC. To further clarify this issue, we also examined the scenario in which the core's EoS is applied throughout the entire mass distribution of the star i.e., the case without a crust, as shown in Fig.~\ref{fig:FG}(d). 
Our findings indicate that similar behavior occurs even when the outer region of the star is modeled using the same EoS as the core. This suggests that SC is influenced not only by the star's internal structure, but primarily by its mass, regardless of the specific distribution of density within it. Nevertheless, further systematic studies are required to fully clarify this behavior.

Additionally, in Fig.~\ref{fig:FG}(d), we present the two analytical solutions considered in this study: (a) the Schwarzschild solution and (b) the Tolman-VII solution. In both cases, the analytical solutions accurately describe the qualitative behavior of the information quantities, with their quantitative aspects also being well-defined within the specified range. These analytical solutions further support the argument that the compact object's mass is the primary factor influencing SC [see also Eqs.~\eqref{eq:schwarzchild_sol_b} and~\eqref{eq:tolman_sol_b}].

%============================================================================================================
\subsection{Quark stars}
%============================================================================================================
An interesting case is that of quark stars, which exhibit a relatively simple structure due to the simplicity of the governing EoS. Figure~\ref{fig:eos_quarks} presents the relevant results for four interaction cases. As quark stars resemble the form of realistic EoSs in the absence of a crust in the mass-radius relation, the behavior of SC is both qualitatively and quantitatively similar to that observed in realistic hadronic EoSs. In this case, the maximum SC occurring at a mass of 0.2 $M_{\odot}$. Once again, the rapid convergence of the four cases for relatively small masses  ($\sim 0.8~M_{\odot}$), despite their differing radii, suggests that SC is primarily determined by mass rather than the object's size.

\begin{figure*}[t!]
\centering
\includegraphics[width=0.9\textwidth]{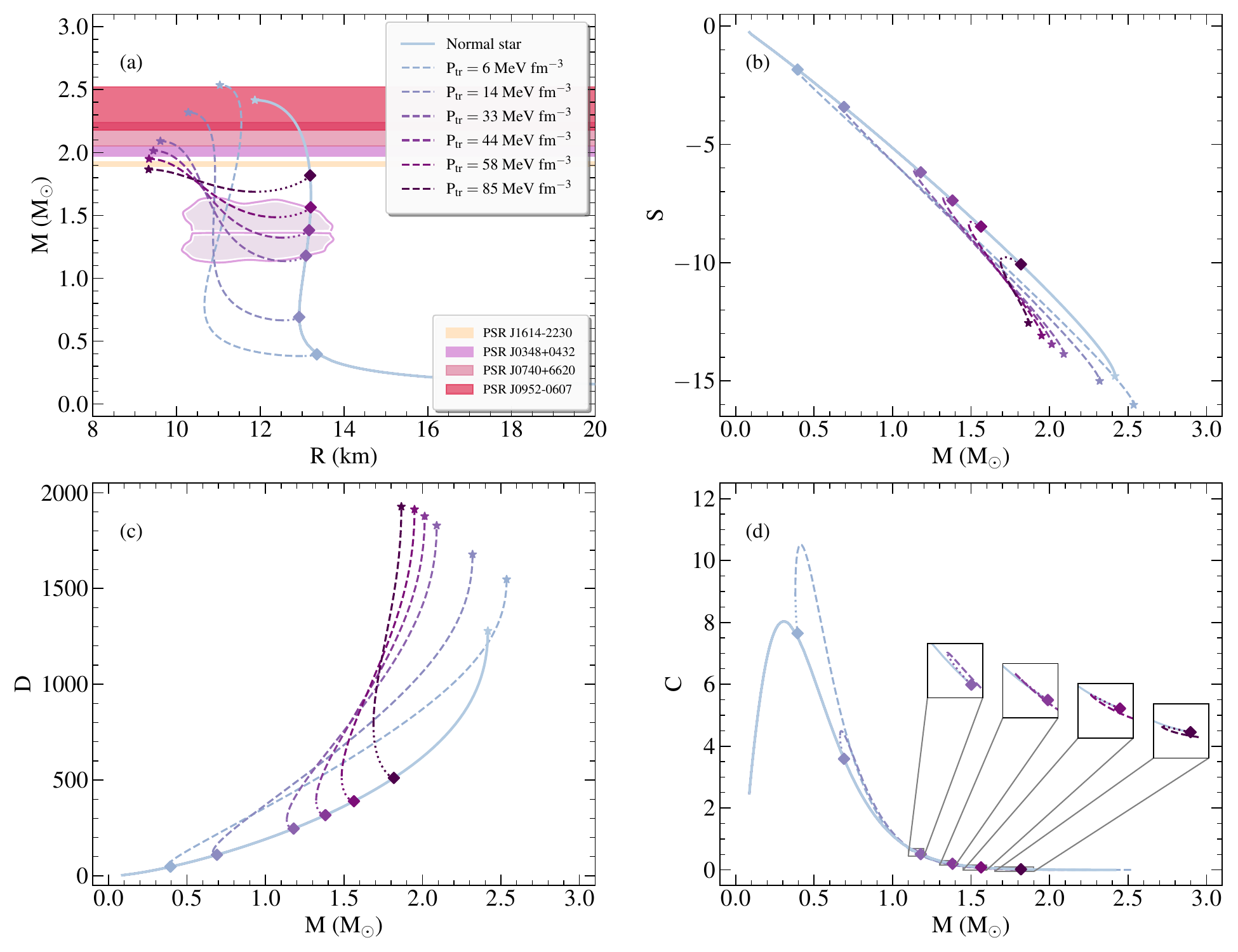}
\caption{(a) Gravitational mass as a function of the radius for a set of hybrid EoSs. The corresponding dependence of (b) the Shannon information entropy, (c) the disequilibrium, and (d) the SC as a function of the gravitational mass. Diamonds indicate the transition points, while stars the stability points. The shaded regions from bottom to top represent the GW170817 event~\cite{Abbott-2019}, and the PSR J1614-2230~\cite{Arzoumanian-2018}, PSR J0348+0432~\cite{Antoniadis-2013}, PSR J0740+6620~\cite{Cromartie-2020}, and PSR J0952-0607~\cite{Romani-2022} pulsar observations with possible maximum neutron star mass.}
\label{fig:mr_hybrid}
\end{figure*}

%============================================================================================================
\subsection{Hybrid stars}
%============================================================================================================
A particularly intriguing case, which has yet to be studied, is that of hybrid stars. Their structural composition is characterized by a phase transition, during which the relationship between density and pressure undergoes a substantial modification, accompanied by a discontinuous jump in energy density. Consequently, it is of significant interest to investigate how such a sudden phase transition within compact objects influences their informational SC. Moreover, identifying the parameters that potentially govern the SC values is essential.

Figure~\ref{fig:mr_hybrid} presents the relevant quantities for six values of transition pressure, $P_{\rm tr}$, with $P_{\rm tr}$ denoting the pressure where the transition occurs. In this context, $P_{\rm tr}$ is considered as the parameter influencing the SC values (analogous results are obtained when using the corresponding transition number density, $n_{\rm tr}$). Across all informational quantities, the hybrid EoS exhibits a markedly different behavior compared to the normal EoS (i.e., one without a phase transition). In all cases, the occurrence of a phase transition introduces a new branch that deviates from the behavior of the normal EoS.

Focusing on the SC measure (though similar trends are observed for all quantities considered), the phase transition results in the appearance of a second peak in SC, the position of which is associated with the value of $P_{\rm tr}$. Notably, as $P_{\rm tr}$ decreases, the distinction in the mass dependence of SC becomes more pronounced. The branch containing the second SC peak resembles the behavior of the normal EoS, suggesting that a strong phase transition in neutron star matter will manifest as a distinct feature in the SC profile. However, as observed in the cases of hadronic and quark stars, SC values rapidly converge beyond $1~M_{\odot}$, rendering the second peak increasingly imperceptible with higher values of $P_{\rm tr}$.

Furthermore, a striking observation emerges: two compact objects with the same gravitational mass can exhibit significantly different SC values. This discrepancy likely arises from the object with higher SC being governed by a more intricate EoS, emphasizing the sensitivity of informational measures to the microphysical properties of dense matter. Therefore, this peculiar behavior observed in SC could be interpreted as a pronounced phase transition impulse occurring within these compact objects.

\begin{figure*}[t!]
\centering
\includegraphics[width=0.9\textwidth]{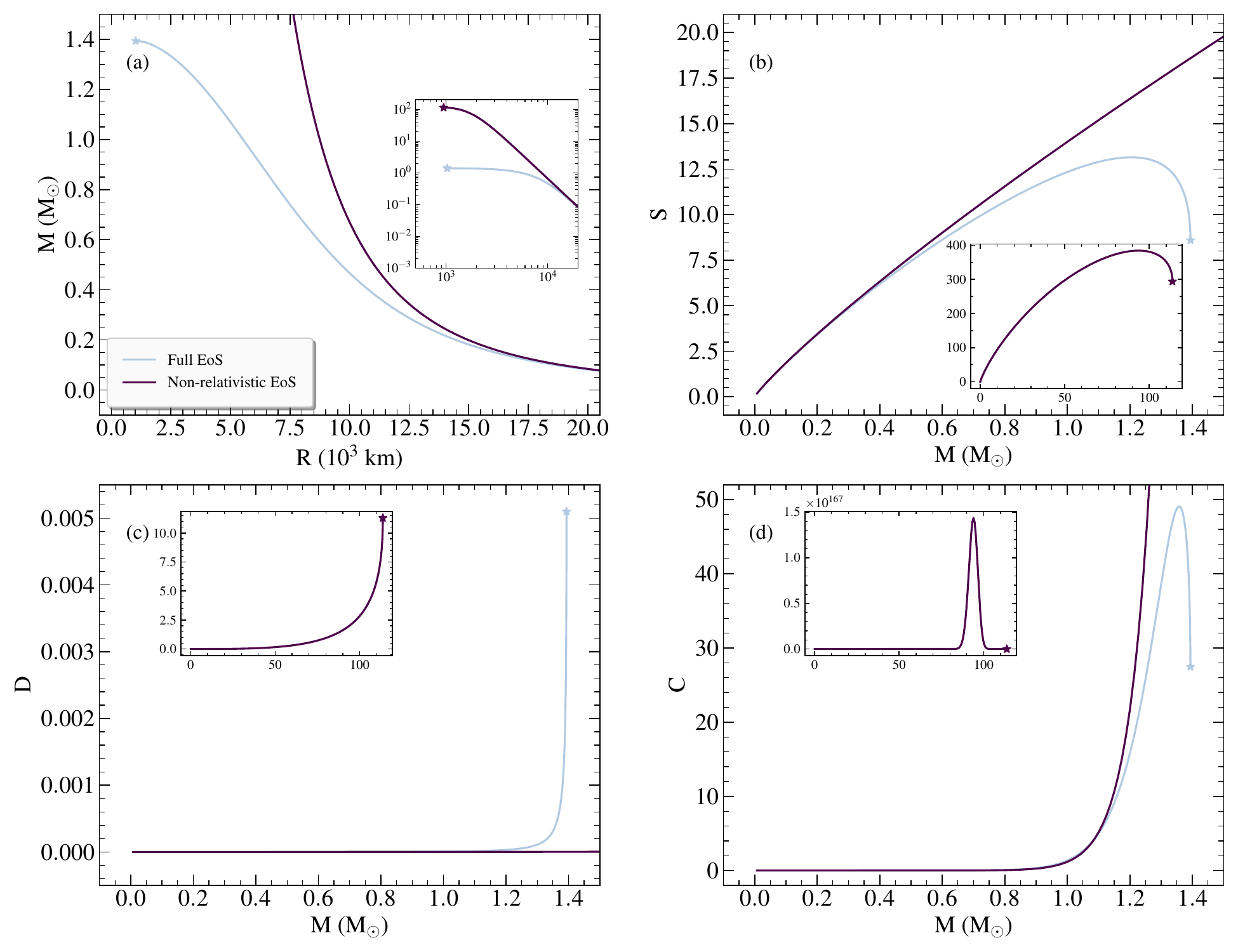}
\caption{(a) Gravitational mass as a function of the radius for white dwarf EoSs. The corresponding dependence of (b) the Shannon information entropy, (c) the disequilibrium, and (d) the SC as a function of the gravitational mass. Stars indicate the stability points.}
\label{fig:dsc_wd}
\end{figure*}

%============================================================================================================
\subsection{White dwarfs}
%============================================================================================================
To provide a comprehensive analysis of compact objects, we also examine the case of white dwarfs. It is worth noting that, in contrast to earlier studies employing the classical Thomas-Fermi method (see for example Ref.~\cite{Sanudo-2009}), the present work utilizes the full set of TOV equations to extract the fundamental properties of white dwarfs.

Specifically, two cases are considered: (a) in the first case, the star is fully described by the EoS defined by Eqs.~\eqref{eq:en_wd}–\eqref{eq:pr_wd}, hereafter referred to as the Full EoS; (b) in the second case, the structure is governed by the non-relativistic EoS given in Eq.~\eqref{eq:wd-non-rel}, denoted as the Non-relativistic EoS. It is important to emphasize that the second case is not physically realistic, as it is valid only at low densities where electrons are non-relativistic and breaks down at higher densities. However, it remains mathematically relevant for several reasons:
(a) it offers an excellent approximation in the regime of low masses and low densities,
(b) it facilitates direct comparison with prior studies (see Ref.~\cite{Sanudo-2009}), where this case was examined within a physically meaningful context, and
(c) it enhances the mathematical completeness of our analysis by illustrating the behavior of the associated information measures.

Figure~\ref{fig:dsc_wd} presents the relevant quantities for both white dwarf EoS models. In case (a), SC increases with mass and reaches a maximum just before the maximum possible mass of white dwarfs ($\sim 1.35~M_{\odot}$), followed by a noticeable decline as the configuration approaches the maximum mass. In case (b), although the qualitative behavior resembles that of the Full EoS, the peak in SC occurs at an unphysical mass of approximately $\sim 95~M_{\odot}$.

We note that the results obtained in this study differ from those reported in Ref.~\cite{Sanudo-2009}. This discrepancy can be attributed to two primary factors: first, Ref.~\cite{Sanudo-2009} adopted a different definition of SC, which plays a critical role in shaping the behavior of information measures, and second, the present study employs the relativistic TOV framework rather than the Thomas-Fermi method to model the structure of white dwarfs~\cite{Membrado_1988}.

%============================================================================================================
\section{Concluding remarks} \label{sec:remarks}
%============================================================================================================
Compact astrophysical objects serve as ideal physical systems for exploring various informational properties linked to their internal structure and dimensions. Their intricate and composite nature, governed by quantum mechanics and shaped by strong gravitational fields, makes them a valuable source of physical and informational insight. In this work, a diverse set of compact objects, including those composed of interacting Fermi and boson gases, neutron stars, quark stars, hybrid stars, and white dwarfs, has been employed to explore an application of information theory to astrophysical systems. Specifically, the information-theoretic quantities, namely Shannon information entropy, disequilibrium, and SC, have been systematically examined as functions of the gravitational mass. This analysis aims to shed light on the correlation between SC and mass, as well as to identify potential signatures of phase transitions within compact stellar configurations.

Across all cases of the various compact objects studied, a common trend emerges: the SC initially exhibits low values at small masses, increases to a peak at intermediate masses, whose exact location depends on the specific nature of the compact object, and subsequently decreases rapidly at higher masses, returning to low values. It is noteworthy that beyond $1~M_{\odot}$, the behavior of the SC for neutron stars, quark stars, and hybrid stars converges, effectively collapsing into a common trend. This convergence suggests a universal behavior in the high-mass regime, regardless of the underlying microphysical description of the compact object. Importantly, this overall behavior is also supported by the analytical solutions considered in this study, further reinforcing its generality. The aforementioned trend suggests that SC is not influenced only by the star's internal structure, but primarily by its mass, regardless of the specific distribution of density within it. This implies that, although the internal composition and phase transitions of a compact object contribute to its overall structure, the gravitational mass predominantly governs its informational characteristics, particularly SC. Such a finding underscores the mass-driven nature of compact objects' behavior and suggests that, at least in certain mass regimes, the specific microphysics of the star may have a secondary effect on the overall SC.

In addition, the results indicate that SC does not exhibit a correlation with the maximum mass values associated with points of extreme stability. In view of previous studies that have demonstrated a connection between CE and these stability points, this observation suggests that SC and CE are not directly related. The findings imply that SC is predominantly governed by the bulk properties of compact objects, while CE is more responsive to the internal density distribution and plays a more significant role in pinpointing the stability point through its minimization.

Furthermore, the role of phase transitions and their signatures in SC has been examined through a twofold approach. In the first scenario, the transition from the outer crustal state to the perfect fluid of the inner core was considered, while in the second, a strong first-order phase transition in the interior of neutron stars to quark matter was investigated. In the former case, a comparison between EoSs with and without a crust, as well as quark matter EoSs, which resemble realistic models without a crust, revealed that the qualitative behavior of SC remains largely unaffected. Only minor quantitative differences were observed. In contrast, the latter case displayed a markedly different behavior: SC developed a distinct new branch and exhibited a second peak before converging to the common trend observed at higher masses. This dependence, which significantly alters both the qualitative and quantitative behavior of SC, highlights its potential as a sensitive indicator of strong phase transitions within compact stars.

Finally, the application of information theory to compact objects, particularly through the study of SC, reveals a qualitatively universal relationship, where the primary contribution stems from the macroscopic properties of these objects, predominantly their mass. This finding underscores the dominance of gravitational mass in shaping the informational characteristics of compact stars, regardless of their internal structure or phase transitions. While this study provides significant insights, it also paves the way for future research that could further explore the interplay between mass, internal structure, and information-theoretic quantities. Future investigations could focus on the role of extreme conditions, such as those found in highly compact objects or those undergoing rapid rotation, and how these factors influence the SC and other information measures. In other words, examining the temporal evolution of compact objects is well suited to exploring dynamic SC. This type of analysis can reveal valuable insights into how certain amounts of information evolve over time and possibly the direction of their changes. Additionally, a deeper understanding of phase transitions, especially the transition to quark matter evident in hybrid stars, could reveal even more intricate connections between astrophysical phenomena and information theory, offering new perspectives on the fundamental nature of compact objects.

%============================================================================================================
\section*{Acknowledgments}
%============================================================================================================
This work is supported by the Croatian Science Foundation under the project number HRZZ-MOBDOL-12-2023-6026 and under the project ”Relativistic Nuclear Many-Body Theory in the Multimessenger Observation Era” (HRZZ-IP-2022-10-7773).

\bibliographystyle{elsarticle-num}
\bibliography{koliogiannis_bib}

\end{document}